\documentstyle[prl,aps,preprint,psfig]{revtex}
\tightenlines
\begin{document}
\title{Point contact spectroscopy of the electron-doped cuprate superconductor
Pr$_{2-x}$Ce$_x$CuO$_4$: The dependence of conductance-voltage
spectra on cerium doping, barrier strength and magnetic field}
\author{M. M. Qazilbash, Amlan Biswas, Y. Dagan, R. A. Ott and R. L. Greene}
\address{Center for Superconductivity Research,
Department of Physics, University of Maryland, College~Park,
MD-20742}

\date{\today}
\maketitle

\begin{abstract}

We present conductance-voltage ($G-V$) data for point contact
junctions between a normal metal and the electron doped cuprate
superconductor Pr$_{2-x}$Ce$_x$CuO$_4$ (PCCO). We observe a zero
bias conductance peak (ZBCP) for the under-doped composition of
this cuprate ($x=0.13$) which is consistent with $d$-wave pairing
symmetry. For optimally-doped ($x=0.15$) and over-doped ($x=0.17$)
PCCO, we find that the $G-V$ characteristics indicate the presence
of an order parameter without nodes. We investigate this further
by obtaining point contact spectroscopy data for different barrier
strengths and as a function of magnetic field.

\end{abstract}

\section{\textbf{Introduction}}

Point contacts between a normal metal and a superconductor have
been extensively employed to probe the quasiparticle states of
superconductors near the Fermi energy. Point contact spectroscopy
(PCS) experiments have also been done to investigate the pairing
symmetry of superconductors and the presence or absence of nodes
in the superconducting order parameter ~\cite{Deutscher1}. In
particular, in the hole doped copper oxide high T$_c$
superconductors (HTSC), PCS and junction tunneling experiments
have shown that the pairing symmetry is predominantly $d$-wave
~\cite{Kashiwayarev}, strongly supporting the results of many
other types of experiments ~\cite{Tsueirev}. Some of the tunneling
experiments have even suggested a doping-dependent order parameter
and the existence of a Quantum Critical Point (QCP) near optimum
doping ~\cite{Dagan1}. In the electron doped (n-doped) HTSC, the
prior PCS and tunneling data has been more controversial and there
is no consensus at present on the implications of the data. The
reason is the absence of a zero bias conductance peak (ZBCP) in
the tunneling spectra ~\cite{Huang,Kashiwaya2,Alff1} which argues
against the presence of $d$-wave pairing symmetry while other
types of experiments ~\cite{Tsuei,Kokales,Prozorov} support a
$d$-wave pairing symmetry. In this paper, we present more detailed
PCS experiments on the n-doped HTSC Pr$_{2-x}$Ce$_x$CuO$_4$ (PCCO)
in an attempt to better understand the pairing symmetry in the
n-doped HTSC at all doping levels.

The theory of quasiparticle tunneling in a point contact junction
between a normal metal and a conventional BCS-type superconductor
has been developed by Blonder, Tinkham and Klapwijk (BTK)
~\cite{BTK}. In the BTK model, the superconductor is assumed to
have an isotropic gap in momentum space. The barrier between the
normal metal and superconductor is modelled by a delta function.
The strength of the delta function is characterized by a
dimensionless parameter $Z$. The $Z=0$ limit signifies a
completely transparent junction while $Z>>1$ is the tunneling
limit. This model has been very successful in explaining the
features of the $G-V$ spectra for tunneling in conventional
$s$-wave superconductors ~\cite{Blonder}. Fig. 1a shows calculated
$G-V$ characteristics for different barrier strengths.

The BTK model has been modified by Tanaka and Kashiwaya
~\cite{Tanaka1} for tunneling in superconductors with $d$-wave
symmetry of the order parameter. The gap in a superconductor with
$d$-wave symmetry is of the form
$\Delta_\textbf{k}$=$\Delta_0$Cos(2$\theta_\textbf{k}$) where
$\theta_\textbf{k}$ is the angle of a wave vector on the Fermi
surface relative to [100]. The order parameter is anisotropic and
changes sign in momentum space. Constructive interference between
the electron-like and hole-like quasiparticles, which experience
different signs of the order parameter, results in formation of
surface bound states at the Fermi energy ~\cite{Hu}. These surface
bound states are also referred to in the literature as Andreev
bound states (ABS) and lead to enhancement of tunneling
conductance at zero bias. The ZBCP is expected for tunneling into
the $ab$-plane for all surfaces except $(100)$ and $(010)$. The
ZBCP reaches maximum height for tunneling into (110) family of
planes and Fig. 1b shows the tunneling spectra for such case. The
$G-V$ tunneling spectra in $d$-wave superconductors are
direction-dependent assuming a specular interface between the
normal metal and superconductor. However, in practice the surface
of the superconductor is rough on the atomic scale and this
results in contribution to the tunneling current from facets of
various orientations. When the effect of surface roughness is
taken into account ~\cite{Fogelstrom}, the $G-V$ spectra for
tunneling into $(100)$ and $(010)$ planes also exhibit a ZBCP.
ZBCPs have been observed in hole-doped HTSC, most notably and
consistently in YBa$_2$Cu$_3$O$_{7-\delta}$ (YBCO) for tunneling
both into $(100)$ and $(110)$ orientations
~\cite{Dagan1,Krupke,Covington}.

The order parameter symmetry may be more complicated than simply
$s$-wave or $d$-wave, especially at the surface. This is of
relevance because PCS essentially probes surface electronic
states. The theory for tunneling into superconductors with more
complex pairing symmetries, for example, anisotropic $s$-wave,
extended $s$-wave and $d+is$ has also been developed
~\cite{Kashiwaya1}.

While most of the cuprate superconductors have holes as their
charge carriers, the family of cuprate superconductors
$R_{2-x}$Ce$_x$CuO$_4$ ($R$ = La, Nd, Pr, Sm, Eu) is unique in the
sense that its carriers are predominantly electrons. Several early
tunneling spectroscopy and PCS experiments were performed on
optimally-doped Nd$_{1.85}$Ce$_{0.15}$CuO$_4$ (NCCO)
~\cite{Huang,Kashiwaya2,Alff1} to investigate the density of
states and pairing symmetry of this compound. All these
experiments revealed a gap-like feature in the $G-V$ spectra of
NCCO with no hint of a ZBCP due to ABS. This led to the conclusion
that the pairing symmetry for NCCO was not $d$-wave but either
conventional or anisotropic $s$-wave, in stark contrast to the
hole-doped HTSCs. However, two recent PCS and Scanning Tunneling
Spectroscopy (STS) experiments have observed ZBCP in
optimally-doped NCCO ~\cite{Hayashi,Mourach}. Recently, we
presented PCS data on under-doped PCCO which provided convincing
evidence for a ZBCP due to ABS ~\cite{Amlan2}. In that work,
evidence for a doping-dependent pairing symmetry was also
presented.

Meanwhile, penetration depth experiments on NCCO also supported
the $s$-wave scenario ~\cite{Dong,Andreone,Schneider}. However, it
was later realized that the magnetic moment of the Nd$^{3+}$ ions
would influence these measurements and might point to a misleading
conclusion. Subsequent experiments on the related optimally-doped
compound, PCCO ~\cite{Kokales,Prozorov}, which is not plagued by
the magnetic moment problem, provide evidence for $d$-wave
symmetry. However, other penetration depth experiments on
optimally-doped PCCO and NCCO ~\cite{Alff2,Skinta1} are more
consistent with $s$-wave pairing symmetry. The work of Skinta {\em
et al.} ~\cite{Skinta1} also presents evidence for a doping
dependence of the pairing symmetry.

Strong evidence of $d$-wave symmetry for optimally-doped NCCO and
PCCO was provided by the phase sensitive tri-crystal grain
boundary junction (GBJ) experiment ~\cite{Tsuei}, angle-resolved
photoemission (ARPES) measurements ~\cite{Armitage,Sato} and Raman
spectroscopy ~\cite{Blumberg}. However, these experiments, like
tunneling and penetration depth measurements, are to some extent
sensitive to the surface of the sample and there is the issue that
the surface may be different from the bulk. Moreover, the doping
near the grain boundaries may be different from that in the bulk
and could influence the results of the tri-crystal GBJ and the GBJ
tunneling experiments. Measurements of bulk properties were
expected to finally resolve the question of the pairing symmetry
but the results have yet to converge. A very small residual linear
term in thermal conductivity measurements of PCCO crystals is
evidence for absence of nodes in the gap and does not lend support
to a pure $d$-wave pairing symmetry ~\cite{Taillefer}. Several
explanations have been put forward for the absence of the residual
linear term including the presence of a complex order parameter
and localization of $d$-wave quasiparticles. The other bulk
measurement, specific heat, argues in favor of dirty $d$-wave
pairing symmetry ~\cite{Hamza}. The term `dirty $d$-wave' refers
to a $d$-wave pairing symmetry in a superconductor which contains
impurities leading to physical properties that deviate from those
of a clean $d$-wave superconductor.

To summarize, despite strong evidence for a $d$-wave symmetry,
there is no consensus on the nature of the pairing symmetry in
optimally doped PCCO and NCCO. The most recent PCS and penetration
depth data favors a doping-dependent pairing symmetry for these
materials: $d$-wave for under-doped and $s$-wave for over-doped.
In this paper we carry out an extensive investigation with PCS on
PCCO thin films below their superconducting transition
temperatures ($T_c$). We study the $G-V$ characteristics as a
function of cerium doping, magnetic field and barrier strength
($Z$).

\section{\textbf{Experimental method}}

PCCO thin films were used for PCS because better control and
homogeneity of cerium is achieved in thin films compared to
ceramic samples and single crystals. We have grown c-axis oriented
PCCO thin films on Strontium Titanate (STO) and Lanthanum
Aluminate (LAO) substrates using pulsed laser deposition. Details
of film growth have been published elsewhere ~\cite{Maiser,Peng}.
Superconducting thin films of PCCO with cerium concentrations of
$0.13$ (under-doped), $0.15$ (optimally-doped) and $0.17$
(over-doped) were grown. The films have a thickness of $2500-3000
\AA$. The films were characterized by resistivity, AC
susceptibility and X-ray diffraction measurements. The T$_c$ for
the under-doped, optimally-doped and over-doped films as
determined from AC susceptibility and resistivity measurements are
in the range $12 \pm 1K$, $21 \pm 0.5 K$ and $14 \pm 1 K$
respectively.

The `spear on anvil' technique is widely used for making the point
contact junction. A sharp metal tip is carefully brought into
contact with the superconductor and the barrier strength is
controlled by varying the pressure of the contact with a screw
mechanism or a transducer (see, for example, the reviews in Ref.
~\cite{Kempen,Yanson}). In some cases, the sharp tip of a
superconducting material has been employed with the normal metal
as the counter electrode ~\cite{Blonder}. However, this method is
difficult to implement in our experiment for the following
reasons. Only c-axis superconducting PCCO thin films can be grown.
Attempts at growing a-axis and $(110)$ oriented superconducting
PCCO films have not been successful. Since we want to tunnel into
the $ab$-plane of this material and the thickness of the films is
only $3000 \AA$, a great deal of difficulty is encountered, not
only in positioning the normal metal tip on the edge of the film,
but also in varying the contact pressure without the tip slipping
off the film edge. Hence, we have devised an alternative technique
for PCS in the $ab$-plane of thin films and have designed and
constructed a new probe for this purpose. We also note that a
variation of this alternative method has been employed
successfully in the past for PCS on superconductors ~\cite{Dolan}.

A schematic side view of our point contact technique is depicted
in Fig. 2. The film is cleaved in air to expose a fresh edge,
electrical contacts on the film are made with Ag-In and it is
glued with a quick-setting epoxy to a sapphire base. The film is
placed in our PCS probe and a Au or Pt-Rh alloy wire of $0.25 mm$
diameter, oriented perpendicular to the plane of the film, is
carefully brought into contact with the cleaved edge of the film.
The point contact junction is formed between the wire and the
sharp edge of the film. The PCS probe is placed in a Helium-4 bath
which can be pumped down to 1.4 $K$. In order to study the effect
of the magnetic field on the point contact spectra and to drive
the PCCO into the normal state, the film is placed with its
ab-plane perpendicular to the magnetic field. In this orientation
the H$_{c2}$ of our optimally-doped PCCO films is 8 $T$ (at 1.4
$K$) and this ensures that the film goes into the normal state at
fields achievable in our laboratory. Therefore, a bevel gear
arrangement is required to change the direction of motion for the
point contact to be made. The Au or Pt-Rh wire is glued to a
sapphire piece which in turn is glued to the threaded `nut' on the
shaft. The bevel gears can be rotated from outside of the Helium-4
bath. When the bevel gears rotate, the `nut' along with the wire
moves towards the edge of the film. The point contact can be made
this way and the pressure of the contact can also be varied. A
high pitch anti-backlash worm gear arrangement at the top part of
the probe outside the Helium-4 bath ensures that the distance
between the wire and the film edge can be controlled to sub-micron
precision. One important advantage of this method over the `spear
on anvil' method is that the point contact is relatively stable
against mechanical perturbations because the wire is also
supported by the edge of the substrate. We make a four probe
measurement to obtain the ($I-V$) characteristics of our junctions
and numerically differentiate it to obtain the conductance. We
also use a standard lock-in technique to obtain the $G-V$ spectra
directly.

\section{\textbf{Results and discussion}}

Since our method of making point contact is unconventional, we
decided to demonstrate its validity and effectiveness by doing PCS
on optimally-doped c-axis oriented YBCO thin films in the same
configuration that we have used for PCS on PCCO. The reasons for
choosing YBCO for this purpose are that YBCO is an HTSC like PCCO
and there is a consensus that the pairing symmetry of
optimally-doped YBCO is predominantly $d$-wave and a ZBCP due to
ABS has been observed in tunnel junctions and point contact
junctions ~\cite{Dagan1,Covington,Wei,Sharoni}. The YBCO films
have a thickness of 2000 $\AA$ and were grown on STO and Neodymium
Gallium Oxide (NGO) substrates. The films have T$_c$ in the range
$89.5 \pm 0.5 K$. Since the $G-V$ spectra for a superconductor
with $d$-wave pairing symmetry are orientation-dependent, the
films were cleaved in different directions. However, we find no
major difference in the features of the $G-V$ spectra for
different cleavage orientations which suggests that surface
roughness at the atomic scale leads to averaging of contributions
from facets oriented in various directions.

Figure 3 shows the $G-V$ characteristics for point contact
tunneling into the ab-plane of YBCO with Pt-Rh alloy wire as the
counter-electrode. One can see from the zero field data that the
width of the ZBCP is 6 meV and the coherence peaks appear at the
gap value of 18 meV. This is qualitatively similar to what is
observed in PCS and Scanning Tunneling Spectroscopy (STS) on YBCO
thin films and crystals ~\cite{Wei,Sharoni} and confirms the
validity and effectiveness of our point contact method.

We also studied the effect of c-axis oriented magnetic field on
the ZBCP as shown in Fig. 3. We saw no splitting of the ZBCP at
low fields unlike what has been reported for tunnel junctions
~\cite{Dagan1,Covington}. We observe that the peak splits
explicitly in a magnetic field of 8 $T$. The origin of the ZBCP
splitting in YBCO is still a matter of debate
~\cite{Dagan1,Covington}. However, direct splitting of the ZBCP in
a magnetic field has not been observed in other tunneling
experiments ~\cite{Alff1,Ekin}. Two of the possible reasons for
the absence of splitting of the ZBCP at low fields are that the
orientation of the films and surface morphologies of the junction
interface are different from those in tunneling experiments in
Ref. ~\cite{Dagan1,Covington}. In our experiment, the magnetic
field is along the c-axis and perpendicular to the plane of film.
Therefore, vortices enter the film even at very low magnetic
fields due to the large demagnetization factor and hence the
Meissner currents become almost field independent. The entry of
vortices into the film at low fields probably marginalizes the
role that Meissner currents are expected to play in the splitting
of the ZBCP. Also, it is theorized that the splitting due to a
sub-dominant order parameter ($is$ or $id_{xy}$) may be suppressed
for rough junction interfaces, especially if the sub-dominant
order parameter is $id_{xy}$ ~\cite{Rainer}. A third reason for
the absence of splitting at low fields in our experiment may be
that the splitting of the ZBCP is expected to occur above a
critical field whose magnitude is proportional to the
transmissivity of the junction ~\cite{Tanakajpn}. Furthermore, the
transmission coefficient decreases exponentially with increasing
barrier strength ~\cite{Wolf}. Since our point contact junction is
more transparent than a tunnel junction, the fields for which we
observe splitting are larger than those observed in experiments
with tunnel junctions.  We also note that the magnitude of the
splitting is 2.1 $meV$ at 8 $T$ in our experiment which is in
fairly good agreement with the magnitude of the indirect splitting
in Ref. ~\cite{Alff1}. The above discussion on the splitting of
the ZBCP in a magnetic field has important implications for our
data on PCCO.

We have performed a systematic and exhaustive study of PCS in the
$ab$-plane of superconducting PCCO thin films of different cerium
concentrations and have observed the $G-V$ spectra for different
barrier strengths and in magnetic fields. We have made point
contacts on films with different nominal cleavage orientations.
More than fifty point contact junctions have been studied.

For PCS in PCCO films with $0.13$ cerium concentration, we observe
a ZBCP at low junction resistances and we have studied the ZBCP in
a magnetic field applied normal to the ab-plane of the film. The
data is shown in Fig 4. The shape and magnitude of the ZBCP
suggest that it originates due to ABS and we have fitted the zero
field data to a calculated $d$-wave $G-V$ spectrum in Fig 4a. The
fit takes into account the temperature dependence of the Fermi
function. Besides the gap value ($\Delta$) and the barrier
strength (Z), the fitting parameters include the quasiparticle
lifetime broadening parameter ($\Gamma$), and the angle between
the normal to the junction interface and the direction of the
maximum gap ($\alpha$). The ZBCP is completely suppressed for
magnetic fields in excess of $H_{c2}$ which is about 4 $T$,
lending further support to the argument that the ZBCP is related
to the superconductivity of the PCCO. Compared to the $G-V$
spectra on YBCO, the ZBCP is not accompanied by a gap-like feature
because of the low barrier strength of this normal metal/PCCO
junction. The magnitude and width of the ZBCP decreases with
increasing magnetic field ~\cite{footnote1} and the ZBCP does not
split. The possible reasons for the absence of splitting of the
ZBCP in PCCO are the same as those advanced earlier for YBCO.
Moreover, the fact that we do not observe splitting of the ZBCP at
all in PCCO may be due to a combination of two phenomena: low Z of
our junction and the low $H_{c2}$ of PCCO compared to YBCO.
Therefore, the absence of splitting of the ZBCP in a magnetic
field in under-doped PCCO is not unexpected.

Fig. 5 shows some representative $G-V$ spectra for PCCO $x=0.13$
for different junction resistances. For a high junction resistance
which implies a higher barrier strength, we observe a gap-like
feature with coherence peaks at the gap value. This can be seen
more clearly in the inset (Fig. 5a). As the junction resistance is
decreased, a ZBCP appears along with the gap-like feature shown in
greater detail in the inset (Fig. 5b). For even lower junction
resistance (and lower Z), a pronounced ZBCP is seen whose width is
the same as the value of the gap as deduced from the fit in Fig.
4a. We have observed ZBCPs for low junction resistances in
approximately 30 percent of our junctions and we believe the ZBCP
appears when the normal metal electrode is able to penetrate the
native surface barrier and make direct contact with the
superconductor. For a junction with high barrier strength in a
$d$-wave superconductor, theory predicts a sharp and pronounced
ZBCP (see Fig. 1b) which we have never observed in our experiments
on high Z junctions.

How do we explain this unusual $Z$-dependence of the $G-V$
spectra? One possibility is the presence of an induced order
parameter at the surface for junctions with low transmission
coefficient (i.e. high Z). This suggestion is based on the paper
by Tanuma {\em et al.} ~\cite{Tanuma} who have proposed that for a
high-Z junction in a $d$-wave superconductor, the $d_{x^2-y^2}$
order parameter is suppressed at the surface and the spectral
weight is transferred to a sub-dominant $is$ or $id_{xy}$ order
parameter. This effect is maximum for tunneling into (110)
orientation of the junction interface. We believe that in the case
of PCCO, the magnitude of this induced order parameter exceeds the
bulk $d_{x^2-y^2}$ order parameter at the surface, which is
plausible according to the theoretical calculations in Ref.
~\cite{Tanuma}. Therefore, we attribute the gap-like feature in
the high-$Z$ limit to an induced order parameter whose magnitude
ranges between 3 $meV$ and 5 $meV$ as estimated from the
separation of the coherence peaks in several of the $G-V$ spectra
(for example, see Fig. 5a). We note that this induced order
parameter is not present in the low-Z junctions and the ZBCP we
observe is due to the $d$-wave nature of the bulk order parameter
whose magnitude from the fit in Fig. 3a is 6 $meV$. Another
possible explanation for the absence of the ZBCP in junctions with
high-$Z$ is that disorder suppresses the ZBCP. This is based on
the results of the experiment on tunnel junctions on YBCO in which
the effect of disorder on the ZBCP was studied ~\cite{Aprili1}.
However, in that experiment, the gap-like feature disappears along
with the ZBCP as disorder increases. Since a pronounced gap-like
feature is seen in our high-$Z$ data, we think it is unlikely that
the absence of a ZBCP is solely due to disorder.

Although the simplest and most likely explanation of the ZBCPs in
under-doped PCCO junctions is the $d$-wave pairing symmetry, we
want to point out that ZBCPs due to ABS are predicted for other
exotic symmetries (e.g. extended $s$-wave, $d+s$ with $d > s$)
~\cite{Tanaka1}. The surface orientation dependence of the $G-V$
spectra predicted for the above order parameters is different for
each one but we are unable to distinguish between these by our
experimental method because our $G-V$ spectra are an average of
contributions from facets of various orientations on the atomic
scale. The existence of roughness at the atomic scale is supported
by our observation that we find no qualitative change in the
spectra for tunneling into different surface orientations.

The point contact spectra for optimally-doped and over-doped PCCO
for different junction resistances are shown in Fig. 6 and Fig. 7
respectively. The $G-V$ features are similar for these two doping
levels and we will discuss these together. For high resistance
junctions we see gap-like features with coherence peaks at the gap
values. This is consistent with previous PCS and tunneling
experiments on optimally-doped NCCO and PCCO
~\cite{Huang,Kashiwaya2,Alff1} where no ZBCP was observed. For
lower junction resistances and hence lower barrier strengths, we
do not observe a ZBCP which suggests the absence of ABS at the
Fermi energy. The shape of $G-V$ spectra are now influenced by
Andreev reflections and qualitatively resemble those of $s$-wave
superconductors in the low/intermediate Z limit (see Fig. 1a). In
general, a dip in the conductance at zero bias in low $Z$
junctions suggests absence of nodes in the order parameter
~\cite{Achsaf}. Therefore, keeping in view the shape of our $G-V$
spectra for low $Z$ junctions on under-doped, optimally-doped and
over-doped PCCO, we propose a transition from a $d$-wave order
parameter for under-doped PCCO to a nodeless gap for optimal and
over-doped PCCO. As an example, we have fitted the $G-V$ data for
the lowest junction resistance on over-doped PCCO to the
generalized BTK model with isotropic gap as shown in Fig. 8a. A
slightly better fit is obtained with a $d+is$ pairing symmetry
calculation, with equal weights of the real and imaginary parts of
the order parameter as shown in Fig. 8b. Both fits include the
effect of thermal broadening. In calculating the best fit to the
data, the lifetime broadening parameter ($\Gamma$) is kept as
small as possible and the angular integral is taken over the full
range $-\pi/2 < \theta < \pi/2$. The purpose of the fits is to
show that in both models the $s$-wave component has the same
magnitude and that the low bias characteristics of the $G-V$
spectra are determined to a large extent by the $s$-wave component
of the order parameter. We also tried a fit to $d+id_{xy}$ pairing
symmetry model with the same constraints on $\Gamma$ and $\theta$
as in the $d+is$ model. However, given the physically reasonable
constraints on $\Gamma$ and $\theta$, the data could not be fit to
the $d+id_{xy}$ model.

Our data points to a transition in pairing symmetry with doping
and this suggests the existence of a Quantum Critical Point (QCP)
at or near optimum doping. For $d$-wave superconductors,
theoretical considerations limit the pairing symmetry transition
across a QCP from $d$-wave to either $d+is$ or $d+id_{xy}$
~\cite{Vojta}. Our low $Z$ data on PCCO is more consistent with a
$d$ to $d+is$ pairing symmetry transition across a QCP near
optimum doping. A recent report also suggests a transition in
pairing symmetry across a QCP near optimum doping in YBCO.
~\cite{Dagan1}.

It has been suggested that the proximity effect for transparent
junctions on $d_{x^2-y^2}$ superconductors leads to an induced
$s$-wave order parameter in the normal electrode
~\cite{Ohashi,Kohen}. This effect is maximum for $(100)$ or
$(010)$ surface orientations and does not occur for $(110)$
orientation. If the absence of a ZBCP in our low $Z$ data on
optimally-doped and over-doped PCCO is attributed to an $s$-wave
order parameter being induced in the normal electrode due to the
proximity effect, then it is difficult to understand why this
effect is not observed in our low $Z$ junctions on under-doped
PCCO. Therefore, we do not think our data can be explained by an
$s$-wave order parameter induced by the proximity effect. Hence,
the $d+is$ order parameter appears to be a property of the surface
of optimally-doped and over-doped PCCO.

One interesting and puzzling aspect is the dependence of the
gap-like features on the barrier strength and this can be seen in
Fig. 6 as well as in Fig. 7. The energy scale of the gap-like
features is equal to half the separation of the coherence peaks or
maxima on either side of zero voltage bias. For over-doped PCCO,
the energy of the gap-like feature in the tunneling limit is 3.2
meV while in the low $Z$ limit we obtain a value of 1.4 meV. For
optimally-doped PCCO, the energy of the gap-like feature in the
tunneling limit is 5 meV while in the low $Z$ limit it is 3.3 meV.
We do not fully understand this phenomenon. However, we attempt to
give a reasonable explanation. It is possible that the effect of
nano-scale roughness at the place of the junction and the extent
of the tunneling cone for quasiparticle injection change with the
barrier strength. This may lead to gap-like features appearing at
somewhat different energies in the spectra for junctions with
different barrier strengths.

We have also studied the dependence of the $G-V$ spectra in the
tunneling limit on the magnetic field applied perpendicular to the
plane of the film (i.e. parallel to the c-axis of the film) and
present our data in Fig. 9a and 9b. Note that the coherence peaks
are completely suppressed at 3 T for optimally-doped PCCO and 2 T
for over-doped PCCO. These fields are less than $H_{c2}$ which is
8 T for optimally-doped PCCO and 4 T for over-doped PCCO. We have
no explanation for this unusual field dependence at the present
time. Notice that the magnitude of the superconducting gap
decreases with increasing magnetic field and it evolves into a
gap-like feature, qualitatively different from the superconducting
gap, at $H \simeq H_{c2}$. We interpret this dip in the density of
states at the Fermi energy as a normal state gap whose origin has
not yet been determined conclusively and which has been discussed
in detail in one of our earlier papers ~\cite{Amlan1}.

Finally, we would like to comment on whether our point contact
junctions are in the ballistic (or Sharvin) limit i.e the contact
radius ($a$) is much less than the mean free path ($l$)
~\cite{Duif}. This is considered an important criteria for the
validity of PCS. The Sharvin formalism was originally developed to
model transport across a micro-constriction between two similar
metals ~\cite{Sharvin}. This formalism was extended by Wexler to
the case where the micro-constriction has a ballistic as well as a
thermal (Maxwell) part ~\cite{Wexler}. Both the Sharvin and Wexler
formulas have been employed in previous works for calculating the
contact radii for point contact junctions between a metal and an
HTSC. However, there is much debate on whether the Sharvin and
Wexler formulas are applicable in their present forms to a point
contact junction between two materials with very different Fermi
liquid parameters and resistivities ~\cite{Gloos,Walti}, as is the
case with a normal metal and a high T$_c$ superconductor.
Moreover, since the high T$_c$ superconductors have low mean free
paths of the order of tens to hundreds of angstroms, the criterion
for point contacts being in the Sharvin limit ($a << l$), as
estimated from the Sharvin and Wexler formalisms, have rarely been
strictly achieved and in most cases the contact radius is either
comparable to or greater than the mean free path in the
superconductor ~\cite{Amlan2,Wei,Gonnelli,D'yachenko}. Authors
have explained this by either considering the less rigorous
condition ($a < l$) as sufficient for a ballistic contact, or
suggesting that this leads to a higher effective $Z$, or
postulating the existence of several parallel ballistic contacts.
Whatever the case may be, there is some agreement that if the
point contact is dominated by the thermal part, its conductance
will decrease appreciably at higher voltage bias ($eV > \Delta$)
due to localized heating ~\cite{Blonder,Duif,Gonnelli,Hasselbach}.
In our $G-V$ spectra, the conductance at higher voltage bias is
either flat or increasing with voltage. Thus, we conclude that our
point contact junctions are not in the Maxwell regime and the
quasi-particle transport through the junctions is predominantly
ballistic. This is a sufficient condition for energy-resolved
spectroscopy of point contact junctions.

\section{\textbf{Conclusion}}

For {\em low barrier strength junctions} on under-doped PCCO, we
observe a ZBCP and show that it is due to $d$-wave pairing
symmetry. We do not observe a ZBCP in low $Z$ PCS data on
optimally-doped and over-doped PCCO and this indicates absence of
nodes in the gap. Our data, along with theoretical considerations,
indicates a $d+is$ pairing symmetry for optimally-doped and
over-doped PCCO. We find that our data for low barrier strength
junctions for all doping levels can be best explained by a $d$- to
$d+is$ pairing symmetry transition across optimum doping. This
indicates the presence of a QCP near optimum doping for PCCO.

We observe gap-like features for {\em high barrier strength
junctions} for all doping levels. For high Z junctions on
under-doped PCCO, we believe that the $d$-wave order parameter is
suppressed at the surface and the spectral weight is transferred
to an $is$ order parameter which leads to gapped spectra. For high
$Z$ junctions on optimally-doped and over-doped PCCO, the gap-like
features are due to a fully gapped Fermi surface. The data on low
$Z$ junctions on over-doped PCCO suggests that a $d+is$ order
parameter leads to this nodeless and fully gapped Fermi surface.
Indeed, data from all the previous PCS and tunneling experiments
on NCCO and PCCO, which consistently showed a gap-like feature can
be explained by an induced $is$ order parameter at the surface of
under-doped PCCO, and by an order parameter without nodes for
optimally-doped and over-doped PCCO.

We have studied the $G-V$ spectra in magnetic fields up to and
greater than $H_{c2}$ of PCCO. We find that the ZBCP in the $G-V$
spectra of under-doped PCCO does not split in magnetic fields up
to $H_{c2}$. This may be due to one or more of the following
reasons: immediate penetration of vortices into our films; surface
roughness of our junctions; the high transmissivity of our
junctions. For $H > H_{c2}$, we observe a normal state gap at the
Fermi energy in the $G-V$ spectra in the tunneling limit for
optimally-doped and over-doped PCCO. The normal state gap is of
the same order of magnitude as the superconducting gap but differs
qualitatively from it. The normal state gap is at least an order
of magnitude less than the pseudo-gap that has been observed in
other experiments on NCCO ~\cite{Armitage2,Onose}.

The authors acknowledge useful discussions with C. J. Lobb, S. M.
Anlage and P. Fournier. We would like to thank the following
colleagues: J. S. Higgins and M. C. Sullivan for assistance with
PCCO and YBCO film growth, and Hamza Balci and Z. Y. Li for some
of the PCCO ceramic pellets. This work was supported in part by
NSF DMR 01-02350.

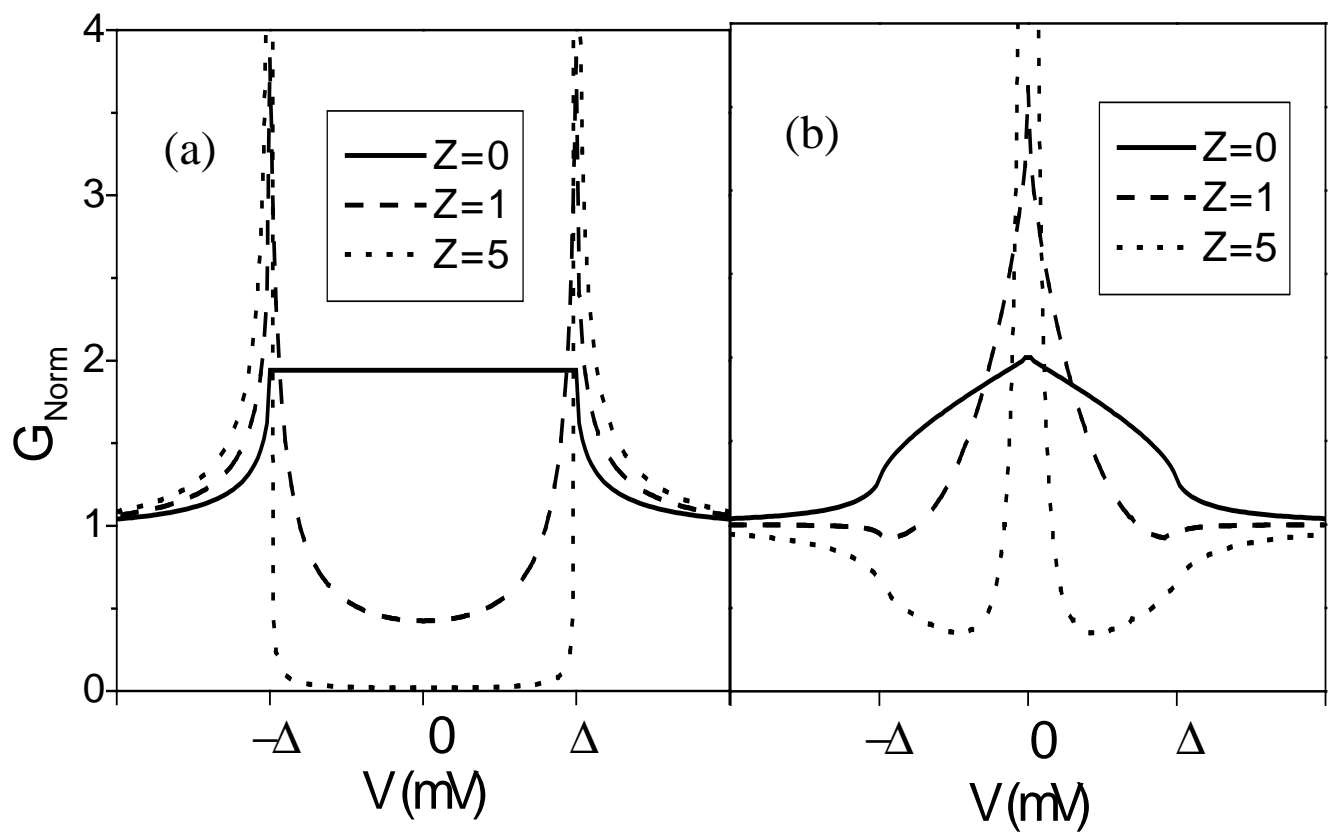
\begin{figure}
\caption{(a) Calculated $G-V$ curves for an $s$-wave
superconductor at zero Kelvin using the BTK model (Ref. [11]). (b)
Calculated $G-V$ curves for a $d$-wave superconductor at zero
Kelvin for tunneling into (110) orientation (Ref. [13]).}
\end{figure}

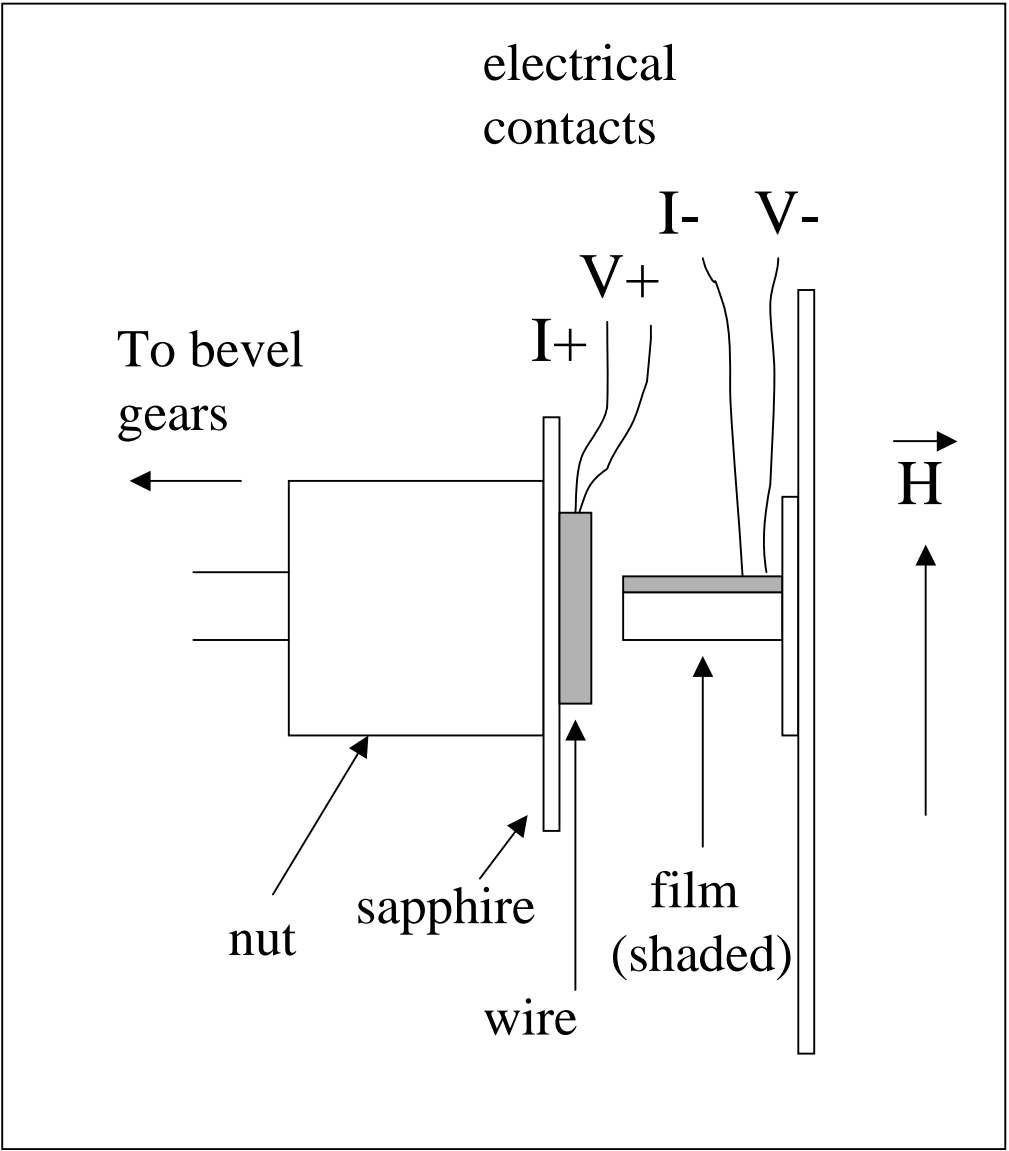
\begin{figure}
\caption{ A schematic side view of our point contact method.}
\end{figure}

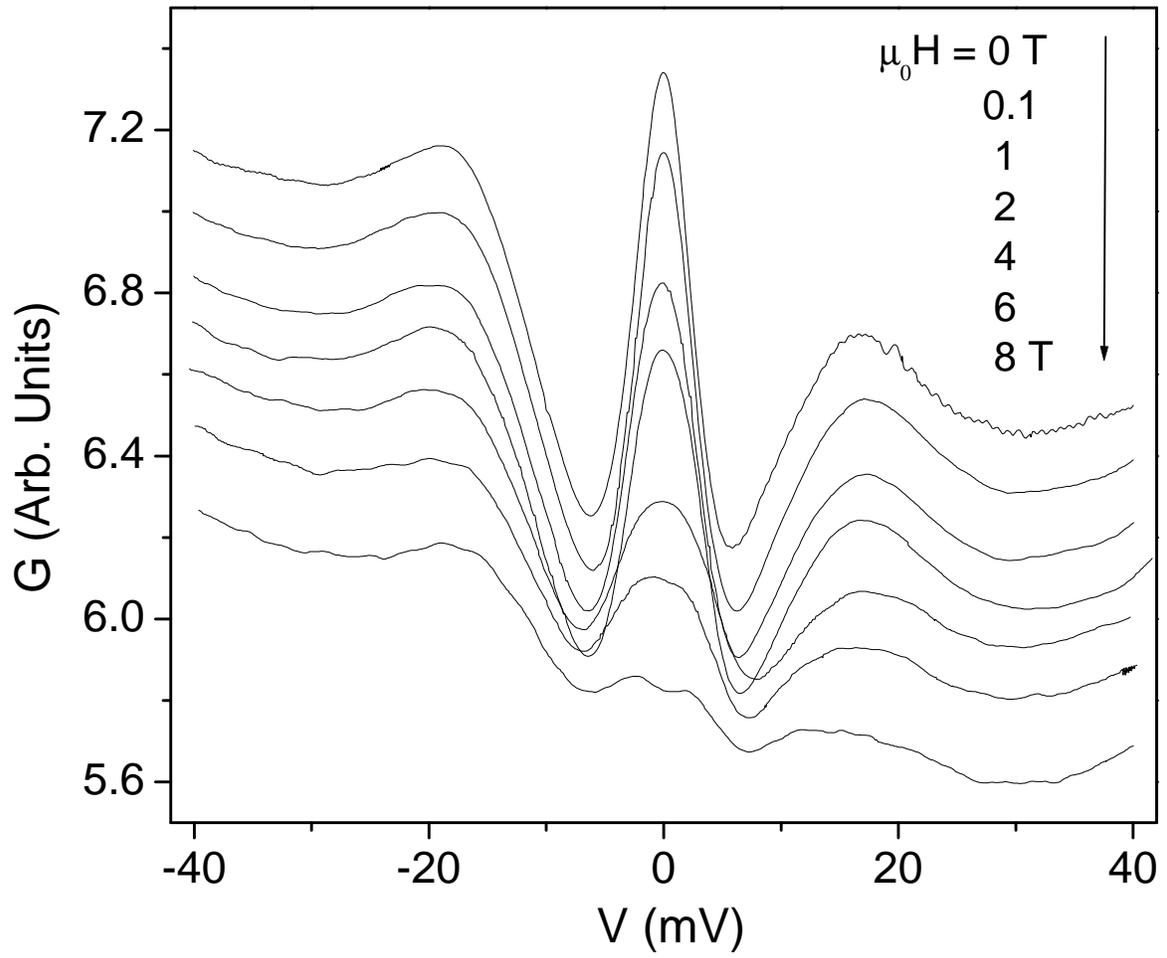
\begin{figure}
\caption{ $G-V$ characteristics for optimally-doped YBCO taken at
T=4.23 K in a magnetic field applied parallel to the c-axis. The
spectra are shifted for clarity.}
\end{figure}

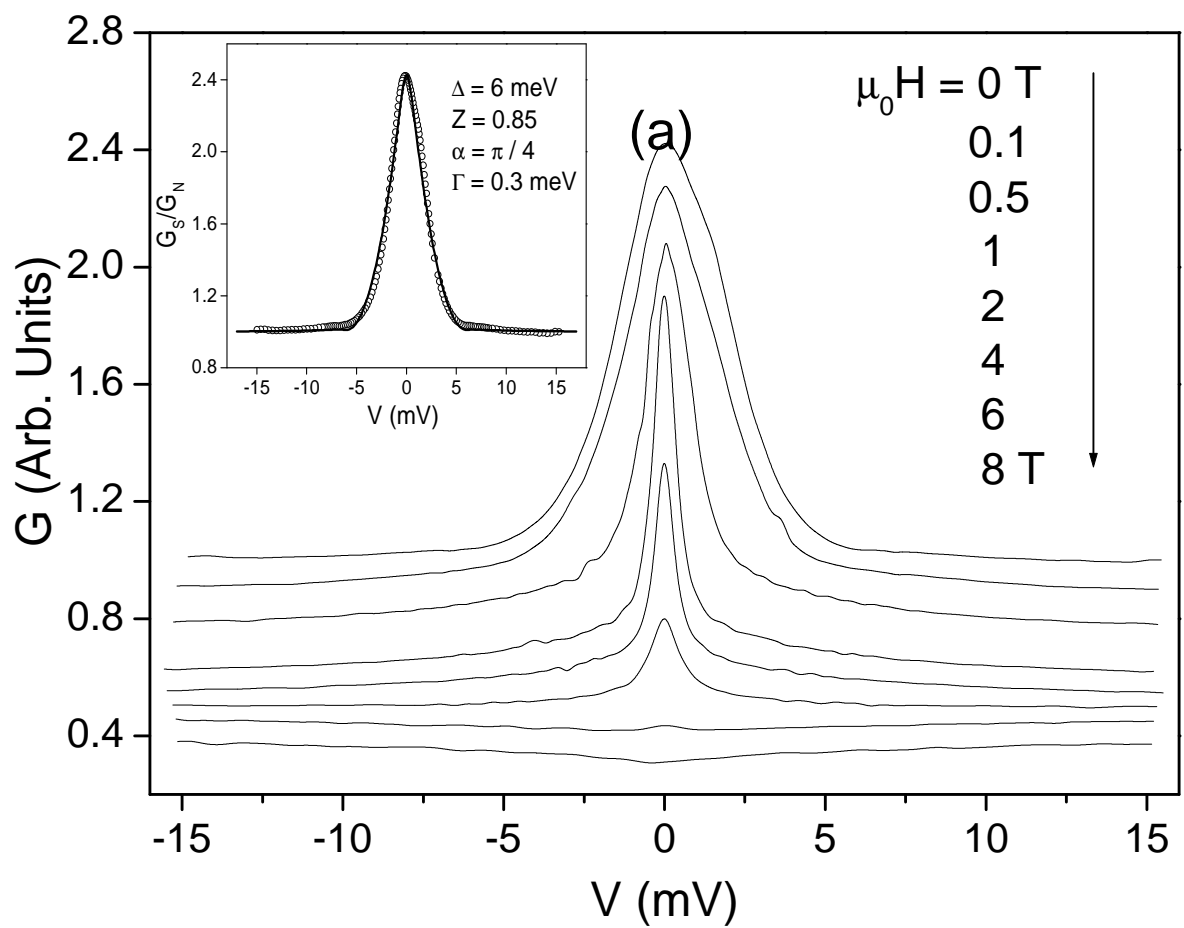
\begin{figure}
\caption{ $G-V$ characteristics for under-doped PCCO at T = 1.43 K
showing the variation of the ZBCP with magnetic field applied
parallel to the c-axis. The spectra are shifted for clarity.
Inset(a): a fit of the ZBCP data (circles) to the calculated
conductance for a normal metal/$d_{x^2-y^2}$ superconductor (solid
line). The fitting parameters are discussed in the text.}
\end{figure}

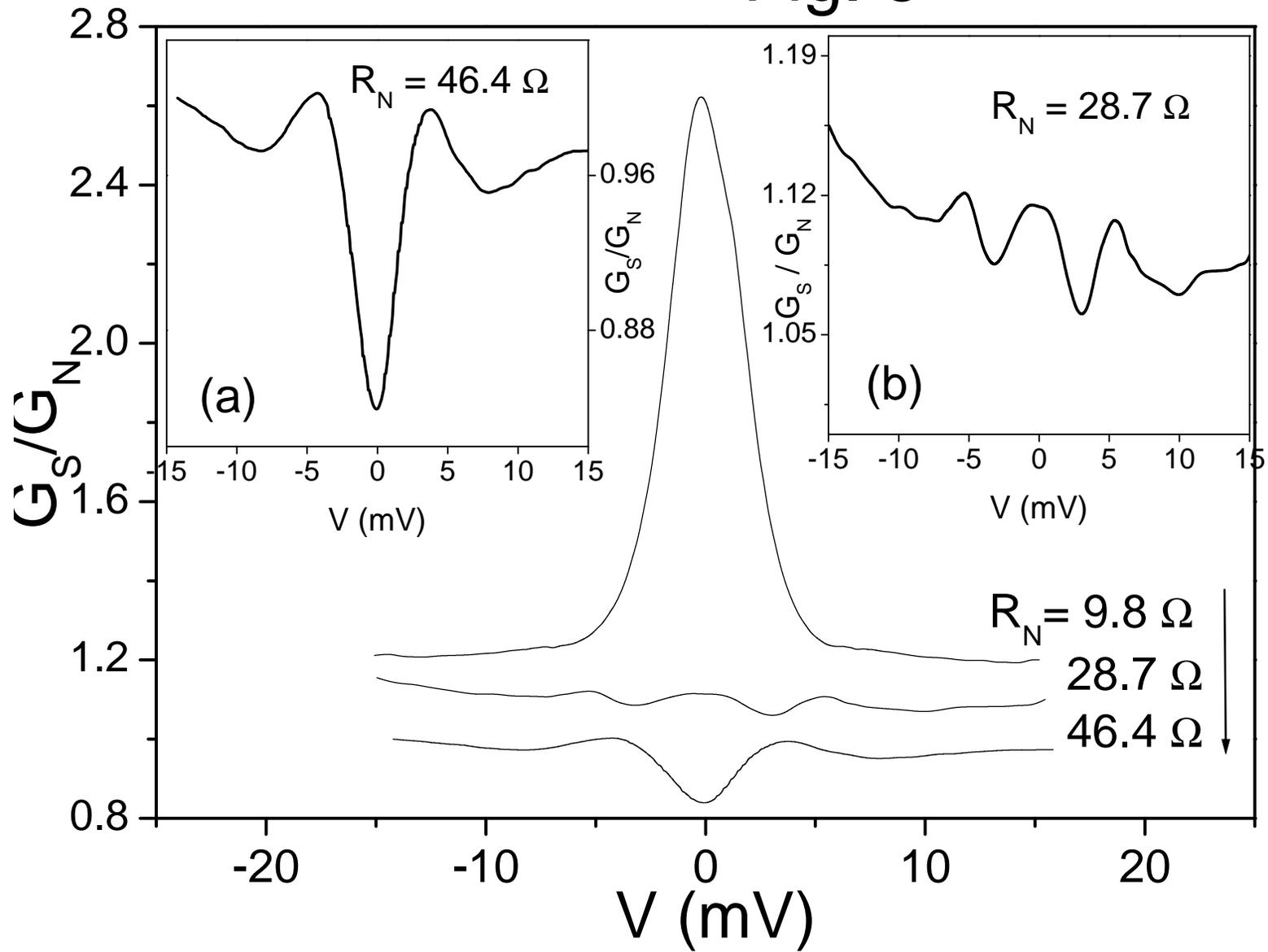
\begin{figure}
\caption{ $G-V$ spectra for different junction resistances for
under-doped PCCO at T = 1.43 K. The conductance is normalized to
the normal state conductance and the spectra are shifted for
clarity. Inset (a) shows in detail the high $Z$ data with $R_N =
46.4 \Omega$. Inset (b) shows in detail the spectrum with normal
state resistance of 28.7 $\Omega$. Note that both the
superconducting gap and ZBCP are clearly visible and compare with
the zero field data for YBCO in Fig. 3.}
\end{figure}

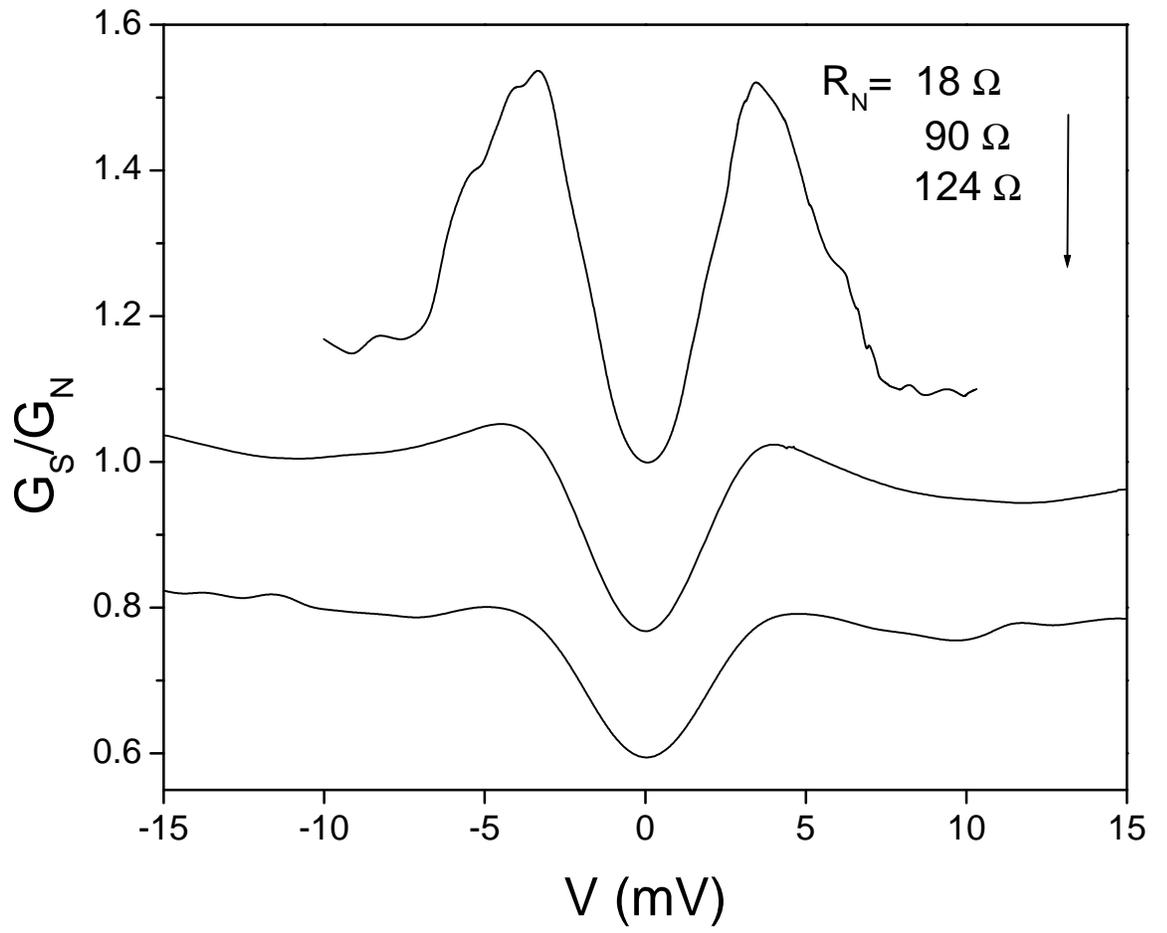
\begin{figure}
\caption{ $G-V$ spectra for different junction resistances for
optimally-doped PCCO at T = 1.43 K. The conductance is normalized
to the normal state conductance and the spectra are shifted for
clarity.}
\end{figure}

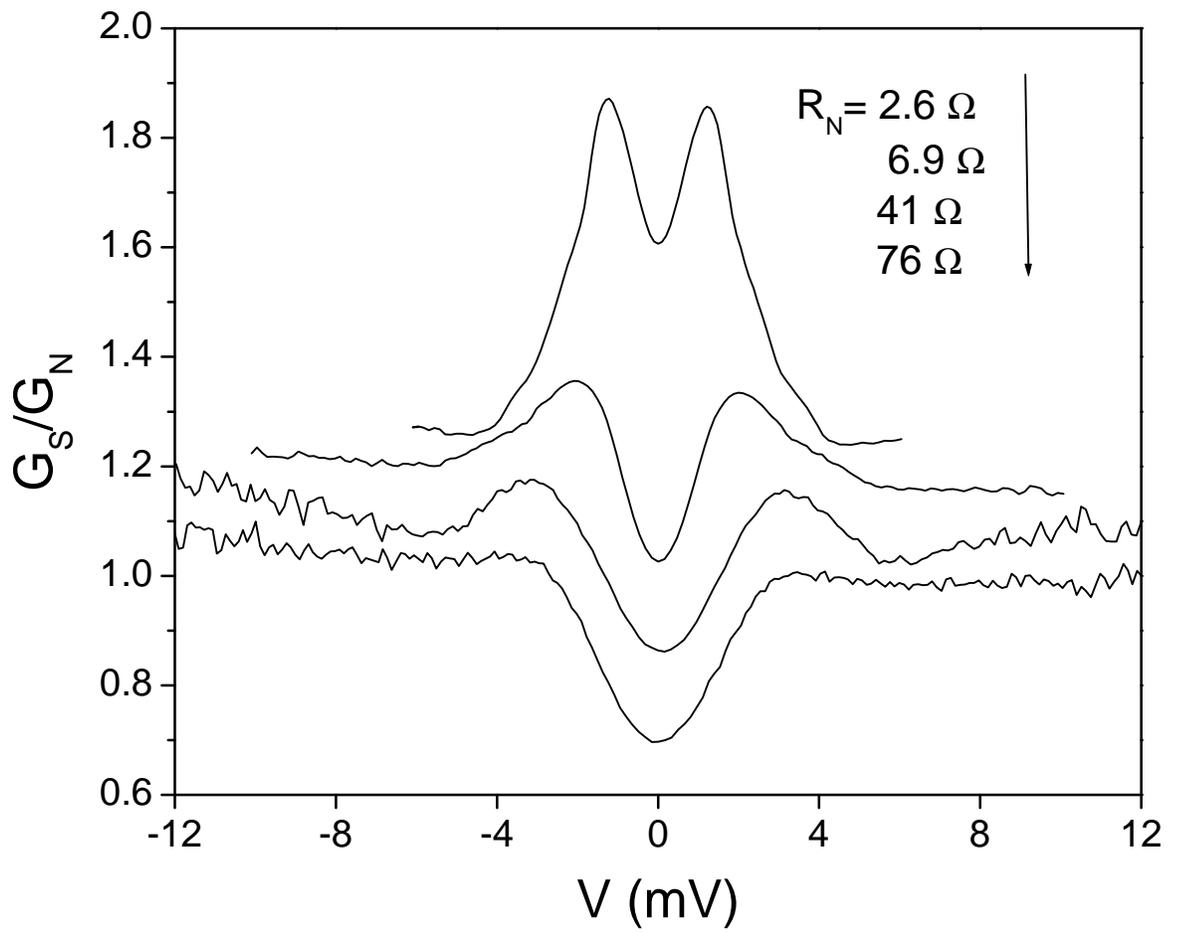
\begin{figure}
\caption{ $G-V$ spectra for different junction resistances for
over-doped PCCO at T = 1.43 K. The conductance is normalized to
the normal state conductance and the spectra are shifted for
clarity.}
\end{figure}

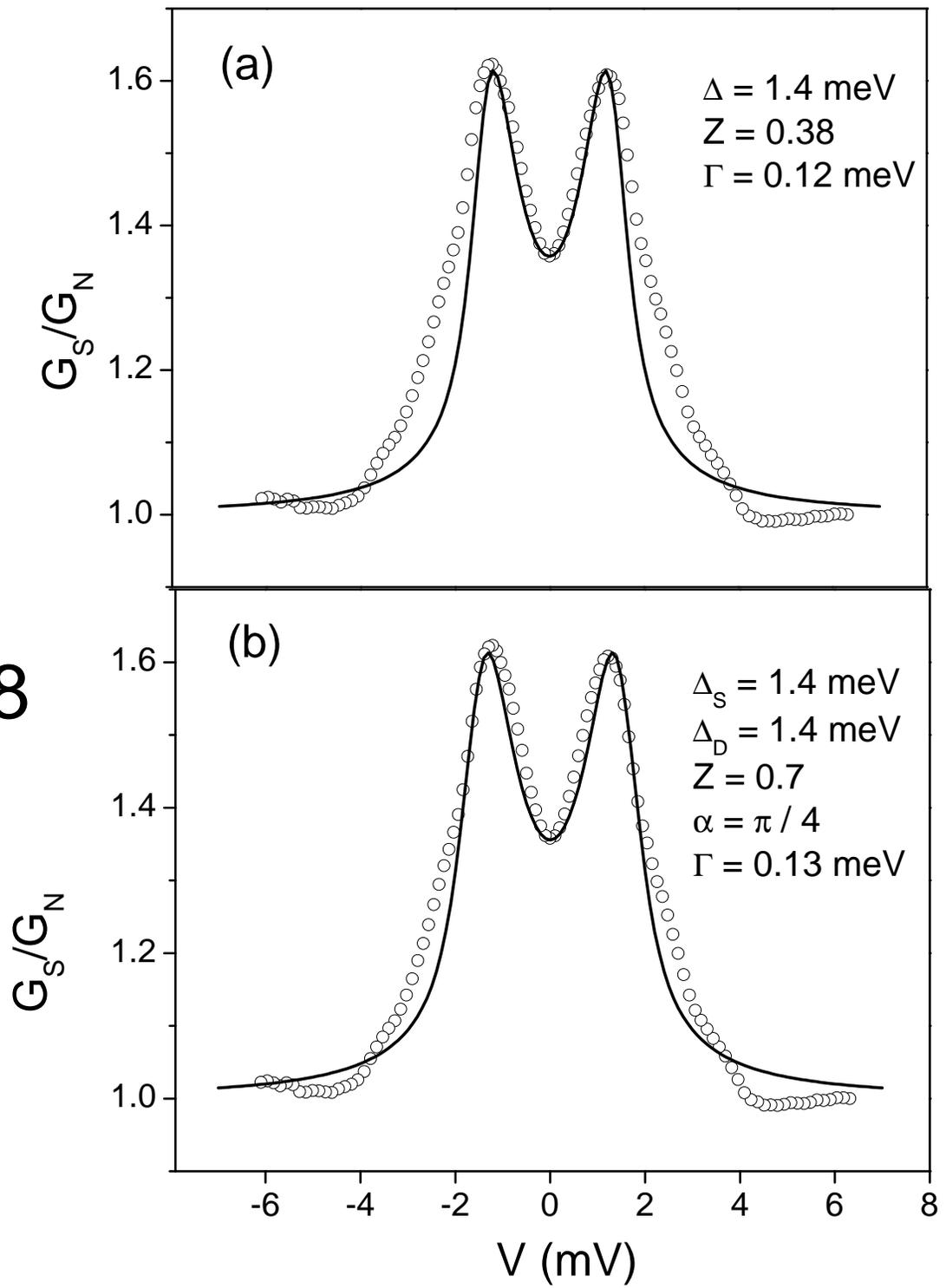
\begin{figure}
\caption{(a) A fit of the low resistance $G-V$ spectrum (circles)
for over-doped PCCO to the generalized BTK model with an isotropic
gap (solid line).(b) A fit of the same data to $d+is$ pairing
symmetry model. The fitting parameters are discussed in the text.}
\end{figure}

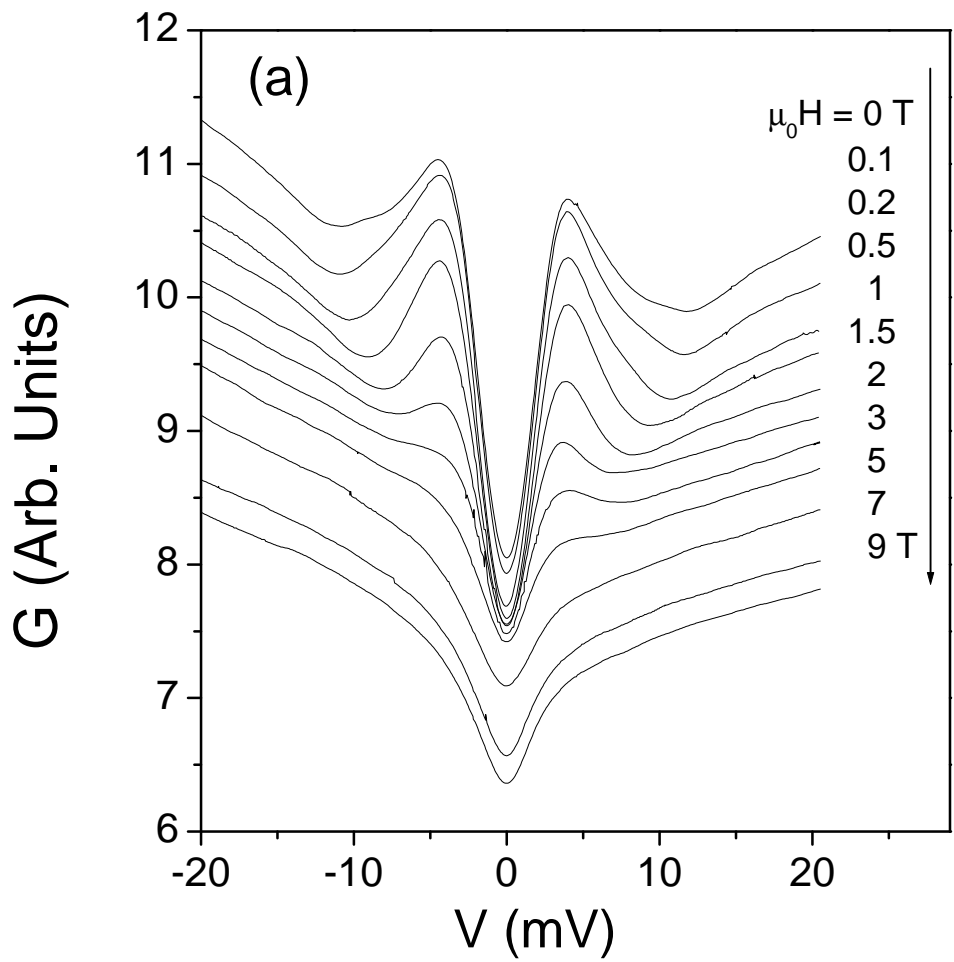
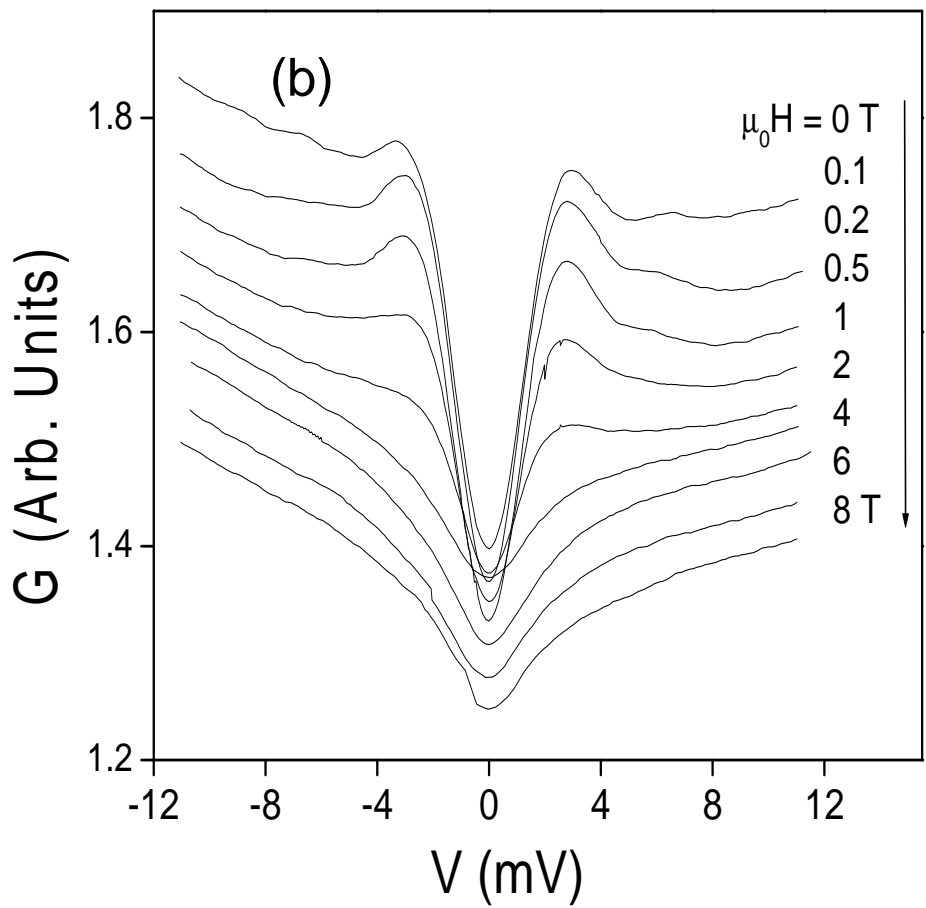
\begin{figure}
\caption{ Variation of the $G-V$ characteristics for high
resistance junctions with magnetic field applied parallel to the
c-axis for (a) optimally-doped PCCO and (b) for over-doped PCCO at
T = 1.43 K. The spectra are shifted for clarity.}
\end{figure}


\begin{references}

\bibitem{Deutscher1} Guy Deutscher and Roger Maynard in: The gap
symmetry and fluctuations in high T$_c$ superconductors, ed.
Julien Bok {\em et al.}, (Plenum Press, New York, 1998)
\bibitem{Kashiwayarev} Satoshi Kashiwaya and Yukio Tanaka, Rep.
Prog. Phys. {\bf 63}, 1641 (2000)
\bibitem{Tsueirev} C. C. Tsuei and J. R. Kirtley, Rev. Mod. Phys.
{\bf 72}, 969 (2000)
\bibitem{Dagan1} Y. Dagan and G. Deutscher, Phys. Rev. Lett. {\bf
87}, 177004 (2001)
\bibitem{Huang} Q. Huang {\em et al.}, Nature {\bf 347}, 369
(1990)
\bibitem{Kashiwaya2} S. Kashiwaya {\em et al.}, Phys. Rev. B {\bf
57}, 8680 (1998)
\bibitem{Alff1} L. Alff {\em et al.}, Eur. Phys. J. B {\bf 5}, 423 (1998)
\bibitem{Tsuei} C. C. Tsuei and J. R. Kirtley, Phys. Rev. Lett.
{\bf 85}, 182 (2000)
\bibitem{Kokales} J. D. Kokales {\em et al.}, Phys. Rev. Lett. {\bf 85},
3696 (2000)
\bibitem{Prozorov} R. Prozorov {\em et al.}, Phys. Rev. Lett. {\bf 85},
3700 (2000)
\bibitem{BTK} G. E. Blonder, M. Tinkham and T. M. Klapwijk, Phys.
Rev. B {\bf 25}, 4515 (1982)
\bibitem{Blonder} G. E. Blonder and M. Tinkham, Phys. Rev. B {\bf
27}, 112 (1983)
\bibitem{Tanaka1} Yukio Tanaka and Satoshi Kashiwaya, Phys. Rev.
Lett. {\bf 74}, 3451 (1995)
\bibitem{Hu} Chia-Ren Hu, Phys. Rev. Lett. {\bf 72}, 1526 (1994)
\bibitem{Fogelstrom} M. Fogelstr\"{o}m {\em et al.}, Phys. Rev. Lett. {\bf 79}, 281 (1997)
\bibitem{Krupke} R. Krupke and G. Deutscher, Phys. Rev. Lett. {\bf
83}, 4634 (1999)
\bibitem{Covington} M. Covington {\em et al.}, Phys. Rev. Lett. {\bf 79},
277 (1997)
\bibitem{Kashiwaya1} Satoshi Kashiwaya {\em et al.}, Phys. Rev. B
{\bf 53}, 2667 (1996)
\bibitem{Ekin} J. W. Ekin {\em et al.}, Phys. Rev. B {\bf 56},
13746 (1997)
\bibitem{Hayashi} F. Hayashi {\em et al.}, J. Phys. Soc. Jpn {\bf 67},
3234(1998)
\bibitem{Mourach} A. Mourachkine, Europhys. Lett. {\bf 50}, 663(2000)
\bibitem{Amlan2} Amlan Biswas {\em et al.}, Phys. Rev. Lett. {\bf
88}, 207004 (2002)
\bibitem{Dong} D. H. Wu {\em et al.}, Phys. Rev. Lett. {\bf 70}, 85(1993)
\bibitem{Andreone} A. Andreone {\em et al.}, Phys. Rev. B {\bf 49},
6392 (1994).
\bibitem{Schneider} C. W. Schneider {\em et al.}, Physica C {\bf 233}, 77 (1994)
\bibitem{Alff2} L .Alff {\em et al.}, Phys. Rev. Lett. {\bf 83},
2644(1999)
\bibitem{Skinta1} John A. Skinta {\em et al.}, Phys. Rev. Lett. {\bf
88}, 207005 (2002)
\bibitem{Armitage} N. P. Armitage {\em et al.}, Phys. Rev. Lett. {\bf 86},
1126(2001)
\bibitem{Sato} T. Sato {\em et al.}, Science {\bf 291}, 1517 (2001)
\bibitem{Blumberg} G. Blumberg {\em et al.}, Phys. Rev. Lett. {\bf
88}, 107002 (2002)
\bibitem{Taillefer} R. W. Hill {\em et al.}, Nature {\bf 414}, 711
(2001)
\bibitem{Hamza} Hamza Balci {\em et al.}, Phys. Rev. B {\bf 66},
174510 (2002)
\bibitem{Maiser} E. Maiser {\em et al.}, Physica C {\bf 297}, 15
(1998)
\bibitem{Peng} J. L. Peng {\em et al.}, Phys. Rev. B {\bf 55},
R6145 (1997)
\bibitem{Kempen} H. Van Kempen in: Scanning Tunneling Microscopy
and Related Methods, ed. R. J. Behm {\em et al.} (Kluwer Academic
Publishers 1990)
\bibitem{Yanson} I. K. Yanson, Sov. J. Low Temp. Phys. {\bf 17},
143 (1991)
\bibitem{Dolan} R. C. Reinertson {\em et al.}, Physica C {\bf
200}, 377 (1992)
\bibitem{Wei} J. Y. T. Wei {\em et al.}, Phys. Rev. Lett. {\bf
81}, 2542 (1998)
\bibitem{Sharoni} A. Sharoni {\em et al.}, Phys. Rev. B {\bf 65}, 134526
(2002)
\bibitem{Rainer} D. Rainer {\em et al.}, J. Phys. Chem. Solids
{\bf 59}, 2040 (1998)
\bibitem{Tanakajpn} Yukio Tanaka {\em et al.} J. Phys. Soc. Jpn, {\bf 71}, 2005 (2002)
\bibitem{Wolf} E. L. Wolf, Principles of electron tunneling
spectroscopy (Oxford University Press, New York, 1985)
\bibitem{footnote1} The decrease in width and height of the ZBCP with
magnetic field is probably due to the decrease in magnitude of the
superconducting gap. This is under further investigation.
\bibitem{Tanuma} Y. Tanuma {\em et al.}, Phys. Rev. B {\bf 64},
214519 (2001)
\bibitem{Aprili1} M. Aprili {\em et al.}, Phys. Rev. B {\bf 57}, R8139 (1998)
\bibitem{Amlan1} Amlan Biswas {\em et al.} Phys. Rev. B {\bf 64},
104519 (2001)
\bibitem{Ohashi} Yoji Ohashi, J. Phys. Soc. Jpn. {\bf 65}, 823
(1996)
\bibitem{Kohen} Amir Kohen and Guy Deutscher, cond-mat/0207382
(2002)
\bibitem{Achsaf} Guy Deutscher {\em et al.}, Physica C {\bf
282-287}, 140 (1997)
\bibitem{Vojta} Matthias Vojta {\em et al.}, Phys. Rev. Lett. {\bf
85}, 4940 (2000)
\bibitem{Duif} A. M. Duif {\em et al}, J. Phys. Condens. Matter
{\bf 1}, 3157 (1989)
\bibitem{Sharvin} Yu. V. Sharvin, Zh. Eksp. Teor. Fiz. {\bf 48},
984 (1965) [Sov. Phys. JETP {\bf 21}, 655 (1965)]
\bibitem{Wexler} G. Wexler, Proc. Phys. Soc. London {\bf 89}, 927 (1966)
\bibitem{Gloos} K. Gloos, Phys. Rev. Lett. {\bf 85}, 5257 (2000)
\bibitem{Walti} Ch. W\"{a}lti {\em et al.}, Phys. Rev. Lett. {\bf 85}, 5258 (2000)
\bibitem{Gonnelli} R. S. Gonnelli {\em et al.}, Eur. Phys. J. B
{\bf 22}, 411 (2001)
\bibitem{D'yachenko} A. I. D'yachenko {\em et al.}, Phys. Rev. B
{\bf 61}, 1500 (2000)
\bibitem{Hasselbach} K. Hasselbach {\em et al.}, Phys. Rev. B
{\bf 46}, 5826 (1992)
\bibitem{Armitage2} N. P. Armitage {\em et al.}, Phys. Rev. Lett.
{\bf 87}, 147003 (2001)
\bibitem{Onose} Y. Onose {\em et al.}, Phys. Rev. Lett. {\bf 87},
217001 (2001)

\end{references}
\end{document}